\documentclass{mem}
\usepackage{natbib}\usepackage{txfonts}\usepackage{balance}
\usepackage{graphicx}
\idline{75}{282}
\begin{document}
\def\teff{$T\rm_{eff }$}
\def\kms{$\mathrm {km s}^{-1}$}

\title{
X-ray Observations of the Chemical Abundances in the Intra-Cluster Medium
}

   \subtitle{}

\author{
H.\ B\"ohringer 
          }

  \offprints{H.\ B\"ohringer}

\institute{
Max-Planck-Institut f\"ur extraterrestrische Physik
Giessenbachstrasse, 85748 Garching
Germany 
\email{hxb@mpe.mpg.de}
}

\authorrunning{B\"ohringer}

\titlerunning{Intracluster Metallicity}

\abstract{Clusters of galaxies as the largest clearly defined objects in our Universe
are ideal laboratories to study the distribution of the most abundant chemical 
elements heavier than hydrogen and helium and the history of their production.
The cluster environment allows us to study the element abundances not only inside
the galaxies, but also in the intergalactic space, the intracluster medium. 
Since the intracluster medium is heated to temperatures of several ten Million 
degrees, we can study the chemical composition of this medium through X-ray
spectroscopy. Up to 13 heavy elements have been detected by X-ray spectroscopy
so far. The element most easily detected in the X-ray spectra is iron.
In massive galaxy clusters we find a larger mass of heavy elements in the 
intracluster medium than in the galaxies. The consideration of the intracluster medium
is therefore vital for an understanding of the complete history of nucleosynthesis of the
heavy elements. The observed abundances for all elements heavier than nitrogen can 
roughly be modeled by using two types of sources: core collapse supernovae
and supernovae type Ia. So called cool-core galaxy clusters show a larger
heavy element abundance in the cluster center which seems to be enriched primarily
by products of supernovae of type Ia. The evidence for observations of an
evolution of the heavy element abundance with redshift has still a moderate significance.

}
\maketitle{}

\section{Introduction}

Clusters of galaxies are the largest clearly defined objects in our Universe.
They comprise masses in the range of about $10^{14}$ to $ 3 \times 10^{15}$
M$_{\odot}$. In the mass range of about few $10^{13}$ to $10^{14}$ M$_{\odot}$
we find the groups and poor clusters of galaxies.
As an integral part of the cosmic large-scale structure clusters form from large-scale
overdense regions in the matter distribution, the seeds of which have
been set presumably at the epoch of inflation. Galaxy clusters form very
late in the history of our Universe and the bulk of the cluster population has 
emerged only after a redshift of 2. Before, hardly any massive clusters existed
but groups of galaxies have been present \citep{boe10,all12}

\begin{figure}[t!]
\resizebox{\hsize}{!}{\includegraphics[clip=true]{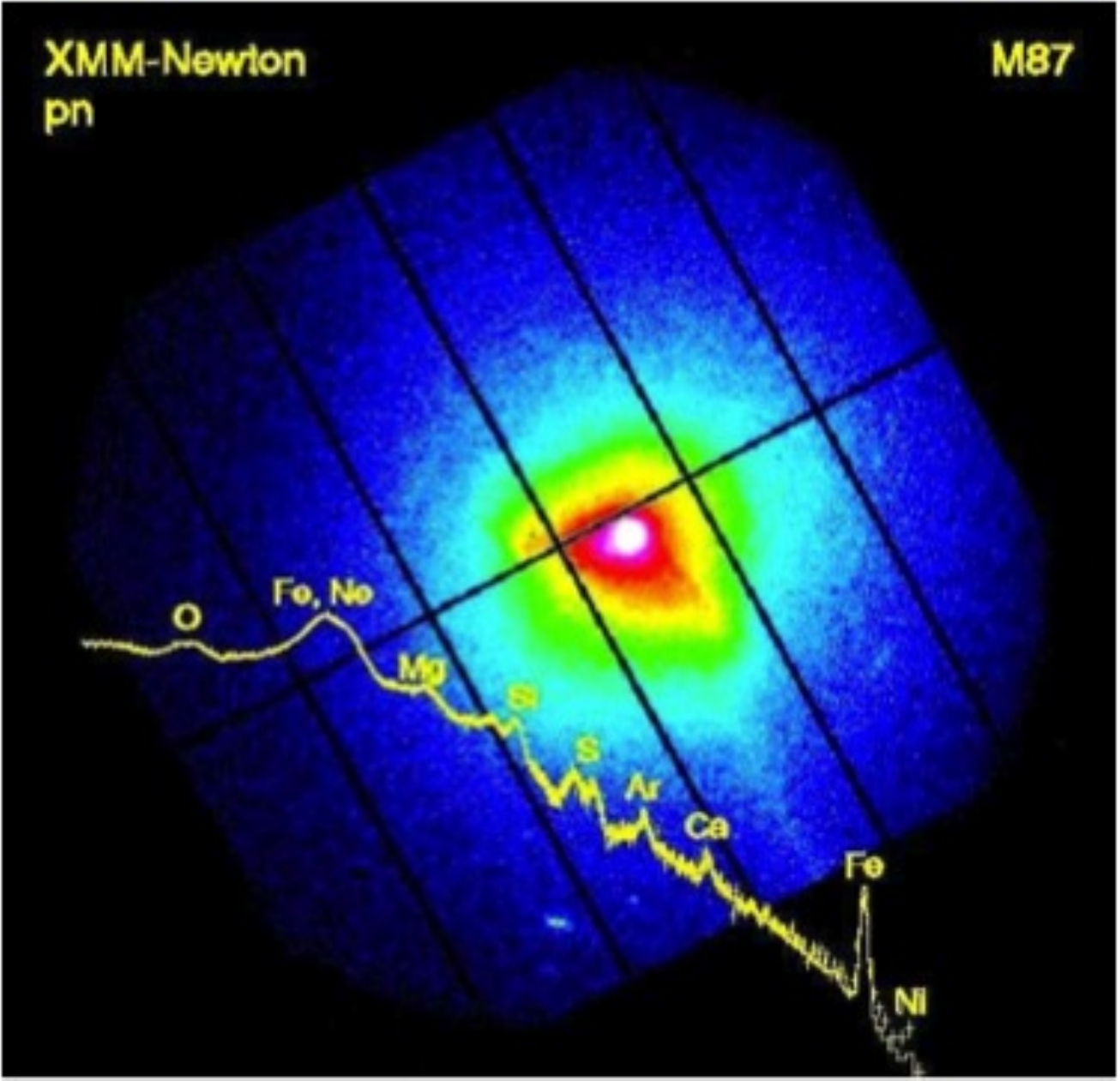}}
\caption{\footnotesize
XMM-Newton image and spectrum of the central region of the X-ray halo
of M87 in the Virgo cluster. The image shows the X-ray surface 
brightness of a region with a size of about 70 kpc radius. Overlayed
is the X-ray spectrum of the inner 20 kpc radius region showing
lines of the most abundant heavy elements O, Mg, Si, S, Ar, Ca, Fe,
and Ni. The X-ray emission represented by the spectrum comes from
the entire volume of the X-ray emitting region seen in the image in
projection.
}
\label{fig1}
\end{figure}

Most of the mass of galaxy clusters, about 84\% in massive systems, is made up by
the so-called Dark Matter, whose nature we still don't know. Only about 4\% is made
up by the stars in galaxies and 12\% by the hot intracluster plasma seen in X-rays.
The formation of a galaxy cluster is thus mostly determined by the gravitational 
dynamics of the dark matter, which forms through gravitational collapse a virialised,
nearly spherical symmetric system, which can approximately be described by e.g. a
NFW model \citep{nfw95}. Galaxies which mostly have been formed before the cluster collapse
and the intergalactic medium are collapsing simultaneously with the Dark Matter.
While the galaxy population gains a velocity dispersion of the order of 1000 km s$^{-1}$
in massive systems, the gas heats up to temperatures of several ten Million degrees
and forms the intracluster medium (ICM). The ICM emits thermal radiation in the form
of soft X-rays, in just the wavelength regime where X-ray telescopes show their
best performance. Thus galaxy clusters are very rewarding objects for imaging
X-ray astronomy \citep{sar86, boe10}.    

This fact, that the cluster ICM can be observed with modern X-ray telescopes, makes
galaxy clusters unique astrophysical laboratories, where we can study all the ``baryonic
matter'', in the galaxies and in the ICM simultaneously in detail.
In the extracluster space, the intergalactic medium is mostly invisible and we
can study it essentially only through absorption effects. A most important finding
comparing ICM and galaxies in clusters is the fact that in massive systems we find 
more mass in heavy elements (metals) in the ICM than in the galaxies. Thus the galaxy
cluster laboratories provide us with the unique opportunity to take a full account of 
all metals produced by stellar nucleosynthesis in a certain, representative  volume 
of the Universe. 

\begin{figure*}[t!]
\resizebox{\hsize}{!}{\includegraphics[clip=true]{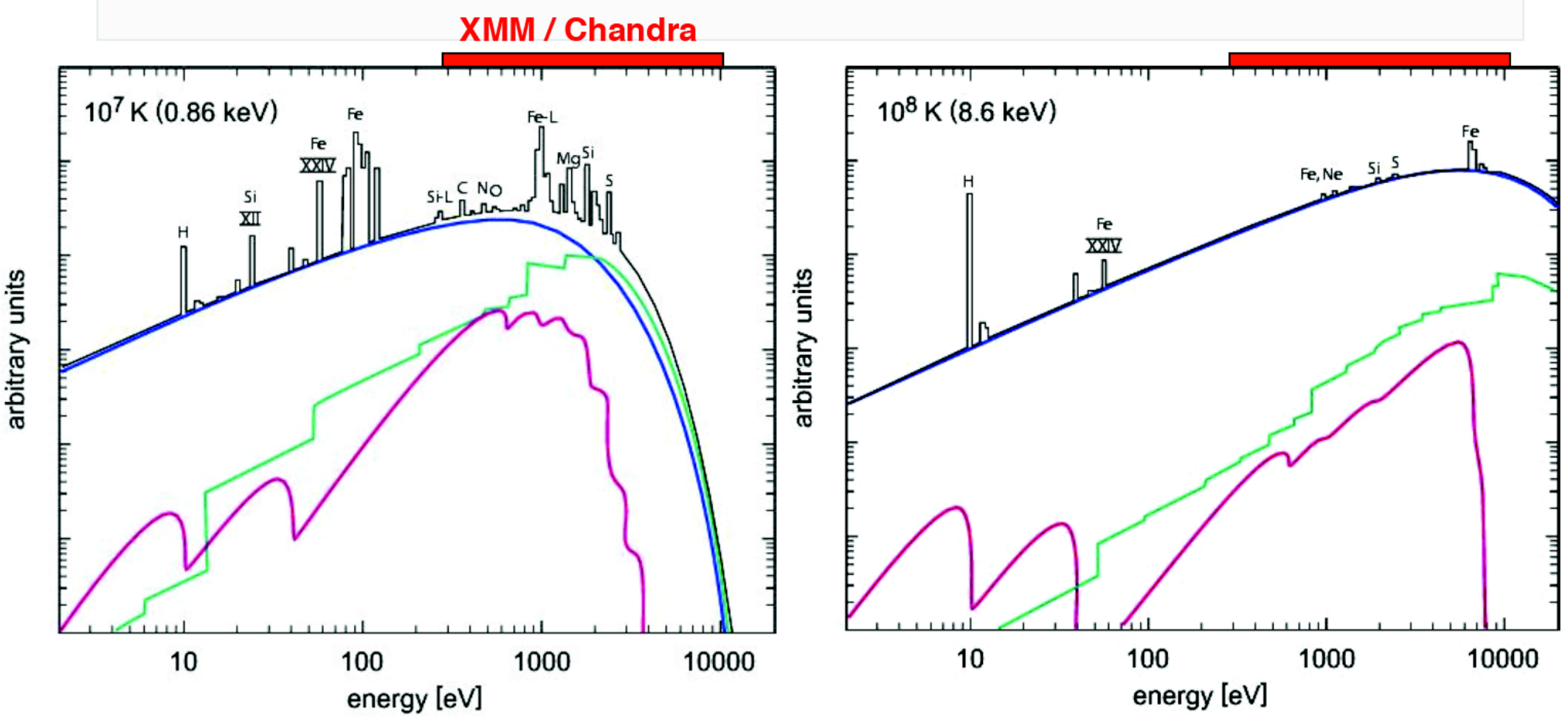}}
\caption{\footnotesize
Theoretically calculated X-ray spectra of hot plasma with solar element abundances
at temperatures of $10^7$ K (left) and $10^8$ K (right). The blue, green, 
and red lines give the contribution of Bremsstrahlung, recombination
continuum, and two-photon emission. Some major lines are labled by the element
responsible for the line emission. The red bar on top of the figure shows
the approximate spectral range covered the XMM-Newton or Chandra detectors
\citep{boe89, bow10}. }
\label{fig2}
\end{figure*}

The X-ray emitting ICM which is heated by shocks and compression during
cluster formation keeps its heat even over a Hubble time, apart from a small
central region in cool core clusters where cooling is more efficient. 
The observed X-ray emission is therefore that of hot plasma effectively in
thermal equilibrium and so far no deviation from thermal equilibrium has been
detected. Therefore the X-ray spectrum is readily modeled as explained in the next section.
X-ray observatories like ESA's XMM-Newton and NASA's Chandra space observatories
detect X-ray photons as single events taking a record of the direction
and energy of incidence. Therefore the photons can be binned in the form 
of images as well as into spectra. Fig. 1 shows as an example the X-ray image and spectrum
of the central region of the X-ray halo of the giant elliptical galaxy M87
in the center of the Virgo cluster \citep{boe01}.

\section{Analysing X-ray Spectra}

The observed X-ray spectra can be interpreted in a straight forward way
due to three major facts. As already mentioned the X-ray emitting plasma
is, as far as we can tell, in thermal equilibrium. Further the plasma
is optically thin, so that practically all the emitted photons in the large 
cluster volume leave the system without being absorbed or scattered (apart
from a few lines in the densest, localized regions in the cluster).
Thus no radiation transfer calculations are necessary
to understand the observations. And finally the plasma is so thin, that 
deexciting collisions are negligible. This means that all excited states
are relaxed by photon emission and the emitted radiation is determined
purely by the collision and excitation rates. 

This implies further that all emission processes originate in a collision
of an electron and an ion: a close collision leads to free-free radiation,
ionizing collisions are followed by recombination radiation, and the
activation of excited states is followed by line radiation or a two photon 
emission continuum in special cases. 
Knowing all the collision and excitation rates as well as radiation
branching ratios one can calculate the emission spectrum, which is essentially
a major bookkeeping exercise, accounting for thousands of atomic processes.
One more aspect that has to be tracked is the distribution of the ionization
stages of each of the considered elements. This is done by considering
the detailed balance of ionization and recombination in thermal equilibrium 
\citep{mew95, bow10}. The two most popular, publicly available radiation 
codes used for the calculation of X-ray spectra of thermal equilibrium plasma
are MEKAL \citep{mew95} and APEC \citep{smi01}.

\begin{figure}[]
\resizebox{\hsize}{!}{\includegraphics[clip=true]{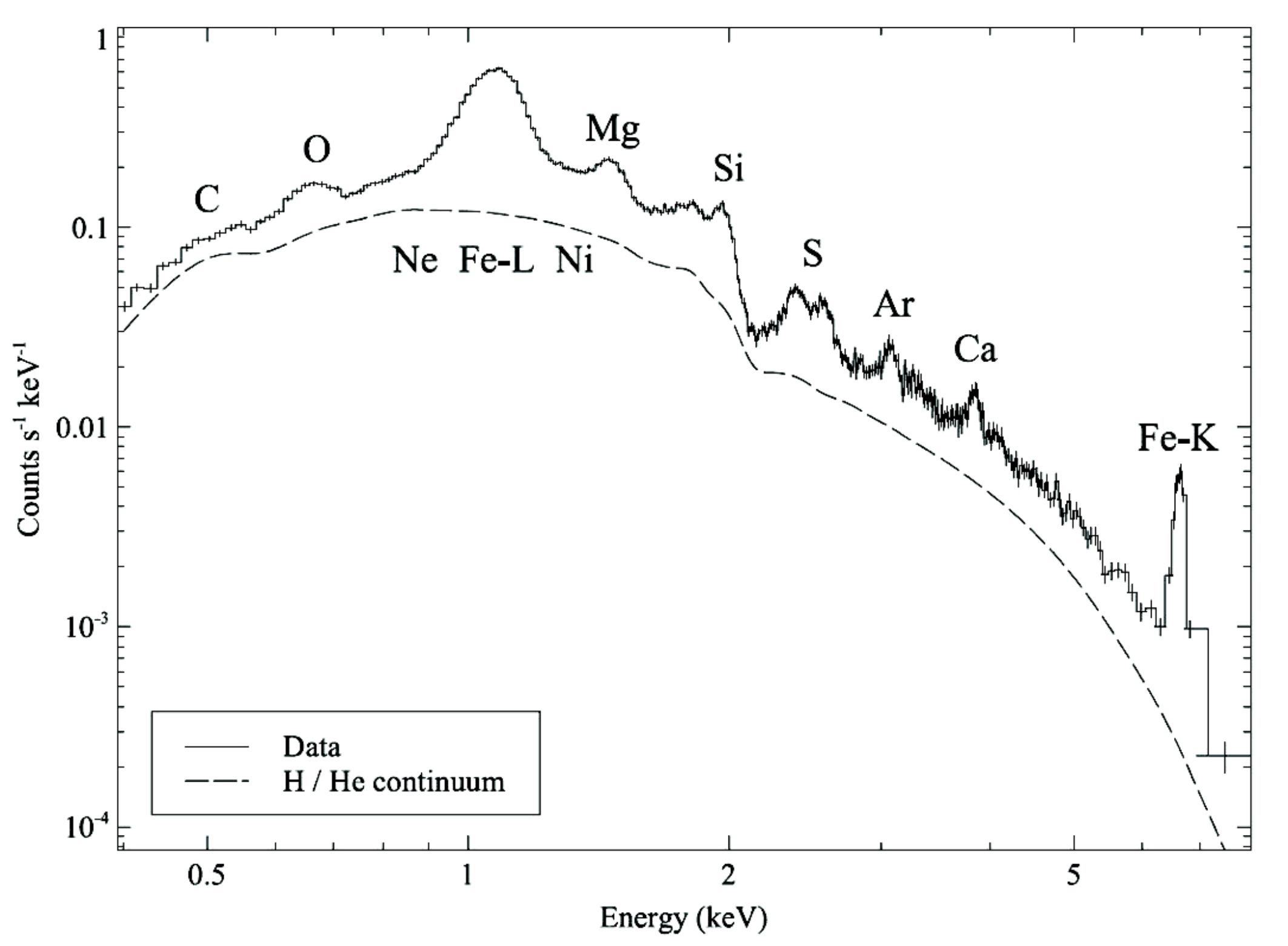}}
\caption{
\footnotesize
X-ray spectrum observed with the XMM-Newton CCD instruments of the ICM of 
the Centaurus cluster \citep{san06}.  
}
\label{fig3}
\end{figure}

Fig. 2 shows typical X-ray spectra for temperatures of $10^7$ and $10^8$
Kelvin, which bracket the typical range of temperatures of the cluster ICM.
At the lower temperature we see more lines from transitions into the K and L
shells. At the higher temperature most of the elements have almost completely lost 
their bound electrons and the spectrum is completely dominated by Bremsstrahlung.
The dependence of all emission on collisions of electrons and ions has the
simplifying consequence that the normalization of the spectrum depends on
the squared plasma density and the form of the spectrum depends on the
temperature and the element abundance.

\begin{figure}[]
\resizebox{\hsize}{!}{\includegraphics[clip=true]{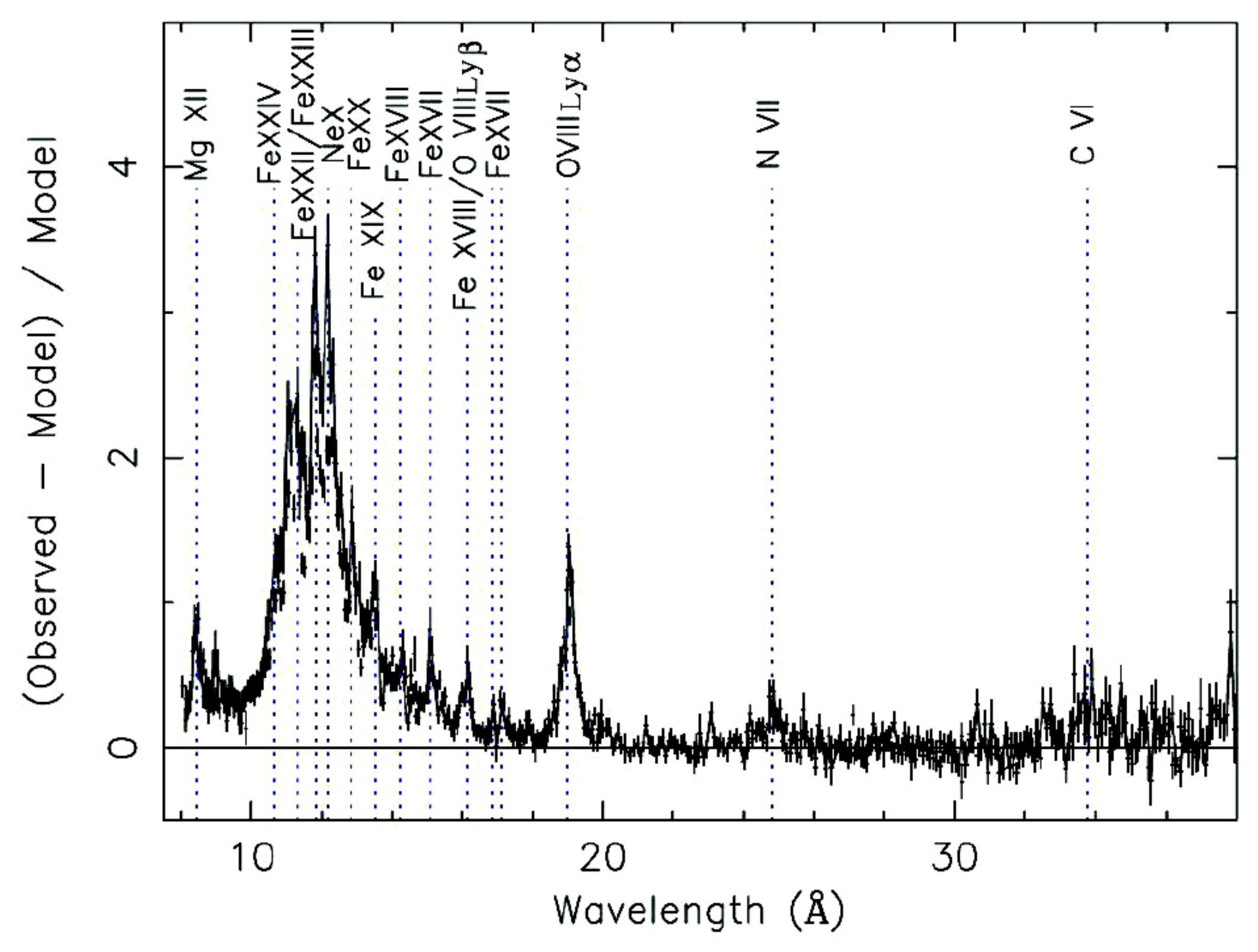}}
\caption{
\footnotesize
XMM-Newton reflection grating spectrometer (RGS) spectrum of the X-ray
halo of M87 \citep{wer06}. The continuum has
been subtracted from the spectrum to enhance the visibility of the
spectral lines.  
}
\label{fig4}
\end{figure}

Fig. \ref{fig3} shows an X-ray spectrum of the cluster ICM of the Centaurus cluster
observed with XMM-Newton, where the spectrum is obtained from the 
energy resolution capability of the X-ray CCD detectors. 
Fig. \ref{fig4} shows the XMM-Newton spectrum produced by the reflection 
grating spectrometer, the XMM-RGS, of the X-ray halo of M87.
The grating spectrometer enables us to resolve the large blend of
lines in the wavelength region 10 - 17 $\AA$. Most of the lines
come from Fe L-shell electrons but there are also lines of Ne
which can only be resolved at this better energy resolution compared
to the CCD spectra.

\section{Significance and origin of the observed abundances} 

\begin{figure}[]
\resizebox{\hsize}{!}{\includegraphics[clip=true]{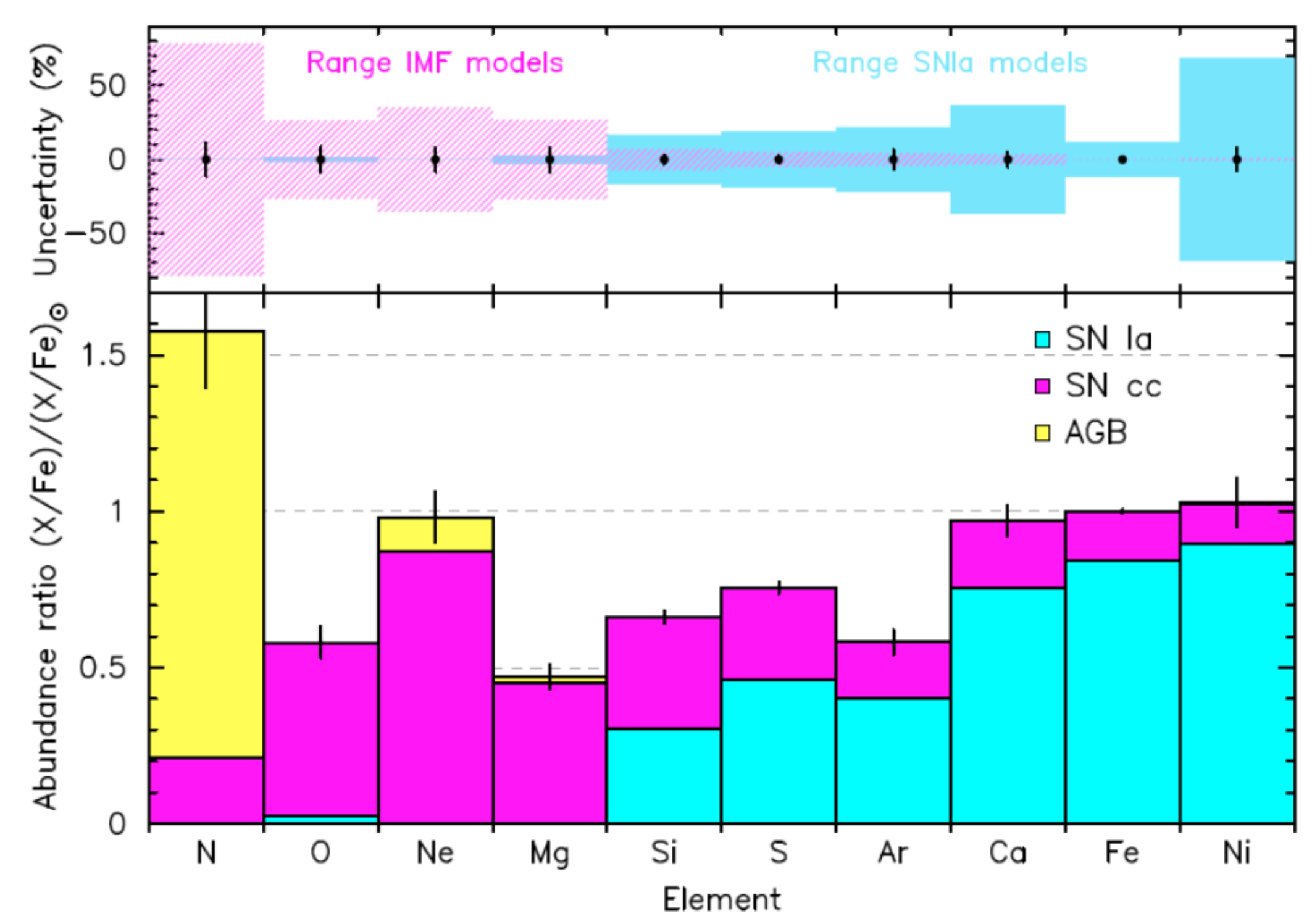}}
\caption{
\footnotesize
Overview over the observed abundances of the 10 most important elements
with respect to the solar abundance \citep{dep12}. The color code of the bars gives
information of their origin from either asymptotic giant branch (AGB) stars,
core collapse supernovae (SN cc) or thermonuclear white dwarf explosions
(SN Ia). The top panel shows the uncertainty of the predicted yields in the
supernova models.
}
\label{fig5}
\end{figure}

\begin{figure}[b!]
\resizebox{\hsize}{!}{\includegraphics[clip=true]{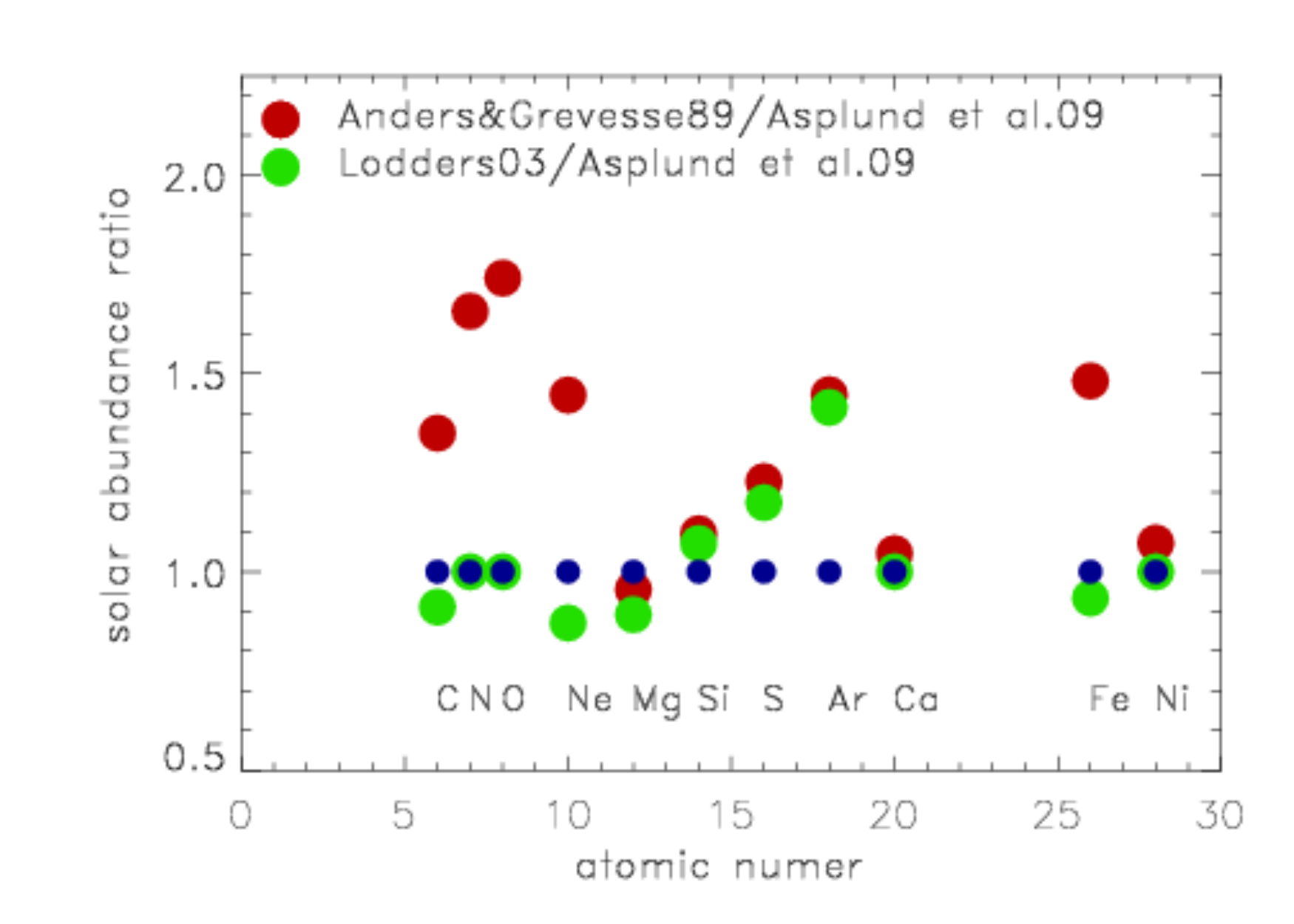}}
\caption{
\footnotesize
Difference in the solar element abundances normalized to those of Asplund
et al. (2009). 
The other two data sets refer to Anders \& Grevesse (1989) and Lodders (2003).
}
\label{fig6}
\end{figure}

11 important elements are visible in the X-ray spectra of deeper observations
of the cluster ICM. Lines of C and N are only observed in the XMM-Newton
RGS spectra, which better cover the lower energies, as can be seen
e.g. in Fig. \ref{fig4}. Ne is only separated from the blend of Fe L-shell lines
in the RGS. The other elements O, Mg, Si, S, Ar, Ca and Ni are best observed
in the CCD spectra of XMM-Newton, Chandra, and Suzaku as shown 
e.g. in Fig. \ref{fig3}. In addition the detection
of Cr and Mn has been reported in the ICM of the Perseus cluster \citep{tam09}.
The most prominent lines in the X-ray spectra are generally those of Fe.
Therefore if metal abundances are reported in the literature from larger
samples of galaxy clusters with not so detailed spectra, these results 
refer primarily to the iron abundance.

\begin{figure*}[]
\resizebox{\hsize}{!}{\includegraphics[clip=true]{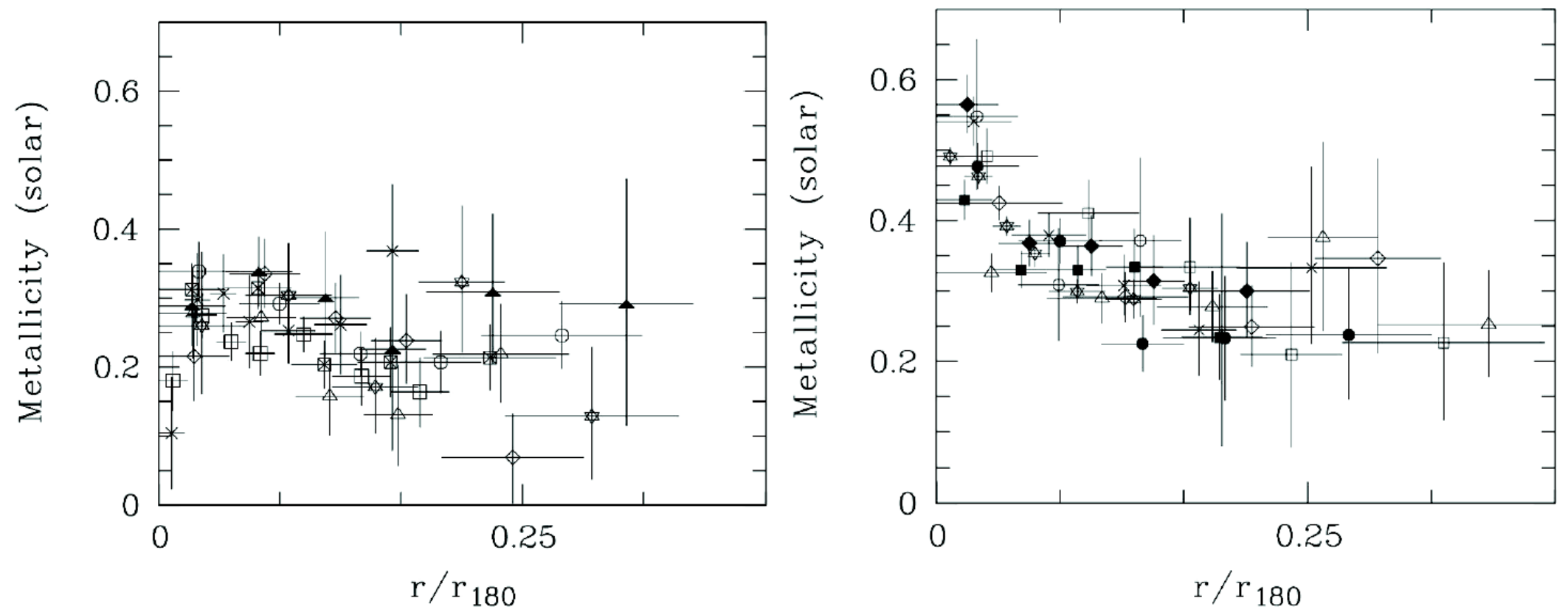}}
\caption{
\footnotesize
Metal abundance profiles in cool core clusters (left) and non-cool
core clusters (right) \citep{dgr01} . The abundances are given relative 
to the solar abundance.
}
\label{fig7}
\end{figure*}

The relative abundance ratio of the elements gives a clue to their origin.
The lightest elements, C and N come primarily from winds of asymptotic giant 
branch (AGB) stars. O and Mg come primarily from core collapse supernovae,
while the primary products of supernovae type Ia (thermonuclear explosions
of white dwarf stars) are Fe and Ni. The intermediate mass elements 
Si, S, Ar, Ca are produced by both types of supernovae. The abundance pattern
of the elements can thus be used for two types of studies. One can reconstruct
what proportion of supernovae of different types is responsible for the 
production of the elements and one can test if the supernovae models
for the production yields of the different elements are consistent with
the observations. Fig. \ref{fig5} illustrates these considerations about the origin 
of the elements, the observed abundance pattern and also uncertainties of the 
model predictions of their production \citep{dep12} . There are still substantial 
uncertainties in the supernova model yields. Therefore the comparison
of the observed abundances with the models gives only an approximate answer,
and the rough consistency between models and observations depends also
on the selected models. A good current account of this situation is given by 
de Plaa et al. (2007).

In the literature the abundances of the elements observed in the cluster 
ICM are usually quoted relative to the solar abundances. This can be the
source of some confusion, since different references are used for the
solar abundances. Most of the literature is still using the work of
Anders \& Grevesse (1989) as reference, which has been updated by a
number of papers e.g. Lodders (2003) and Asplund et al. (2009). Because
this is an important issue for any modeling based on literature data,
we show the difference in these most used references in Fig. \ref{fig6}. 
We note that there are in particular large differences in the very
important diagnostic elements O and Fe. It would be very important 
that the literature converges to a new modern standard.

\section{Spatial distribution of the elements}

\begin{figure*}[t!]
\resizebox{\hsize}{!}{\includegraphics[clip=true]{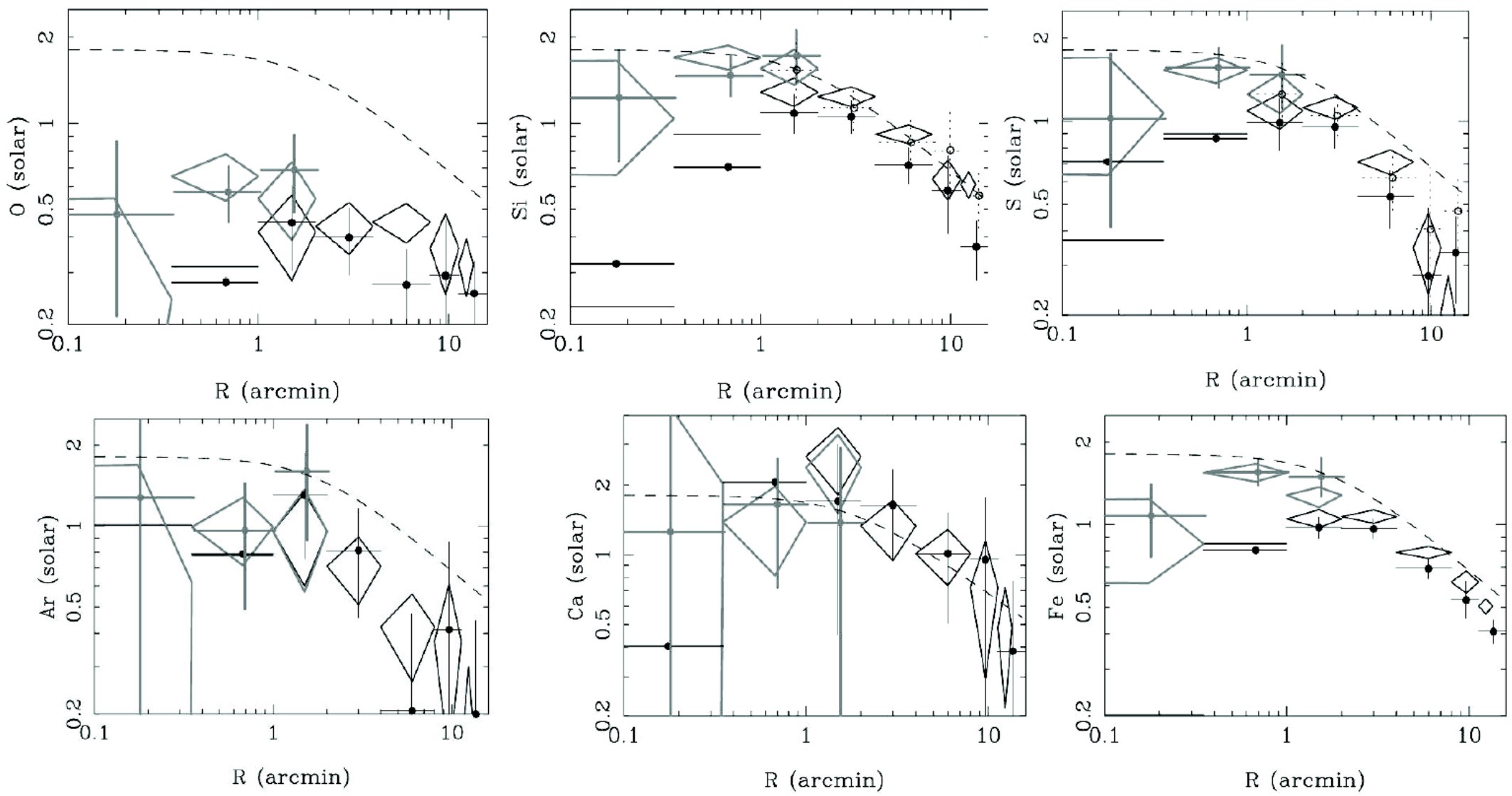}}
\caption{
\footnotesize
Metal abundance profiles of O, Si, S, Ar, Ca, and Fe in the halo of 
M87 observed with the CCD instruments of XMM-Newton \citep{mat03}.
The dashed line is fitted to the Si abundance profile and is repeated
in each panel for comparison.
}
\label{fig8}
\end{figure*}

The Japanese satellite observatory ASCA and the Italian/Dutch
Beppo-SAX carried the first instruments that allowed an effective
simultaneously spatial and spectral resolution of the element
lines. Thus mostly through the Fe line the metal abundance could be
traced in the ICM e.g. in radial profiles. Two families of 
abundance profiles were observed. One family showed a strong 
increase of the metal abundance towards the center  
and it was found that these profiles belong to the group of
cool core clusters. These clusters feature a very dense central 
region in which the cooling time is smaller than the Hubble time.
They also harbour a giant elliptical galaxy in the center,
and feedback effects of the power output of the active nucleus of 
these galaxies prevents  a strong, unimpeded cooling of the 
central ICM. Other clusters with a less dense central region and
often with signatures of recent cluster mergers have much shallower
abundance profiles as shown in Fig \ref{fig7} \citep{dgr01}.

The explanation for this difference comes from the fact that the
cool core clusters have a giant elliptical galaxy which is approximatly
at rest in the center. These galaxies have a rather old stellar
population without massive stars in which no supernova of the core
collapse type has been observed. But they still produce heavy elements
from supernova type Ia, which are rich in iron group elements.
With the large stellar mass concentrated at the center, there is a
continued enrichment of this region by the central galaxy. The
large central excess iron masses observed imply that the center of 
these cool core clusters must have been quiet and undisturbed to
allow this enrichment for almost a whole Hubble time
\citep{boe04}. We also observe that the stellar
mass in the central galaxy is much more concentrated than the 
metal abundance peak, which implies that there are transport
processes which redistribute the metals over a larger central region
(e.g. De Grandi et al. 2004; Rebusco et al. 2006).

With even more spatially resolved spectral details we can look for 
abundance variations as a function of the radius. Studies before
the launch of Chandra and XMM-Newton remained controversial.
The two large X-ray observatories then allowed a major advance.
Fig. \ref{fig8} shows early results for the metal abundance distribution
in the X-ray halo of M87 \citep{mat03}. This central region of
the Virgo cluster is by far the closest region where we can study
the abundance variations in some detail. The Figure shows the 
abundance profiles of the elements O, Si, S, Ar, Ca, and Fe. We
can easily see that the oxygen profile is flatter than all other
profiles. Since O is almost exclusively produced by core collapse 
supernovae (SN cc), this element gives a hint that SN cc products should be
less concentrated in the center. Since Fe on the other hand is produced
primarily by supernovae type Ia (SN Ia), 
which still explode in the old stellar population 
in the cluster center, is seems obvious that this element traces the late
enrichment by SN Ia. The intermediate mass elements of which the 
best observed is Si, would then be expected to have a profile
with a steepness somewhere in between O and Fe. But we see in the figure,
that the Si profile is very similar to that of Fe. 

This makes the interpretation less straight forward, and has led to a some
speculation, e.g. to assume different types of SN Ia contributing to these elements
(e.g. \citep{fin02}). We also have to keep in mind that besides 
the supernovae also stellar winds contribute to the central ICM. While
these stellar winds do not in general carry freshly synthezised material,
they still contribute to the enrichment of the ICM.

\begin{figure}[]
\resizebox{\hsize}{!}{\includegraphics[clip=true]{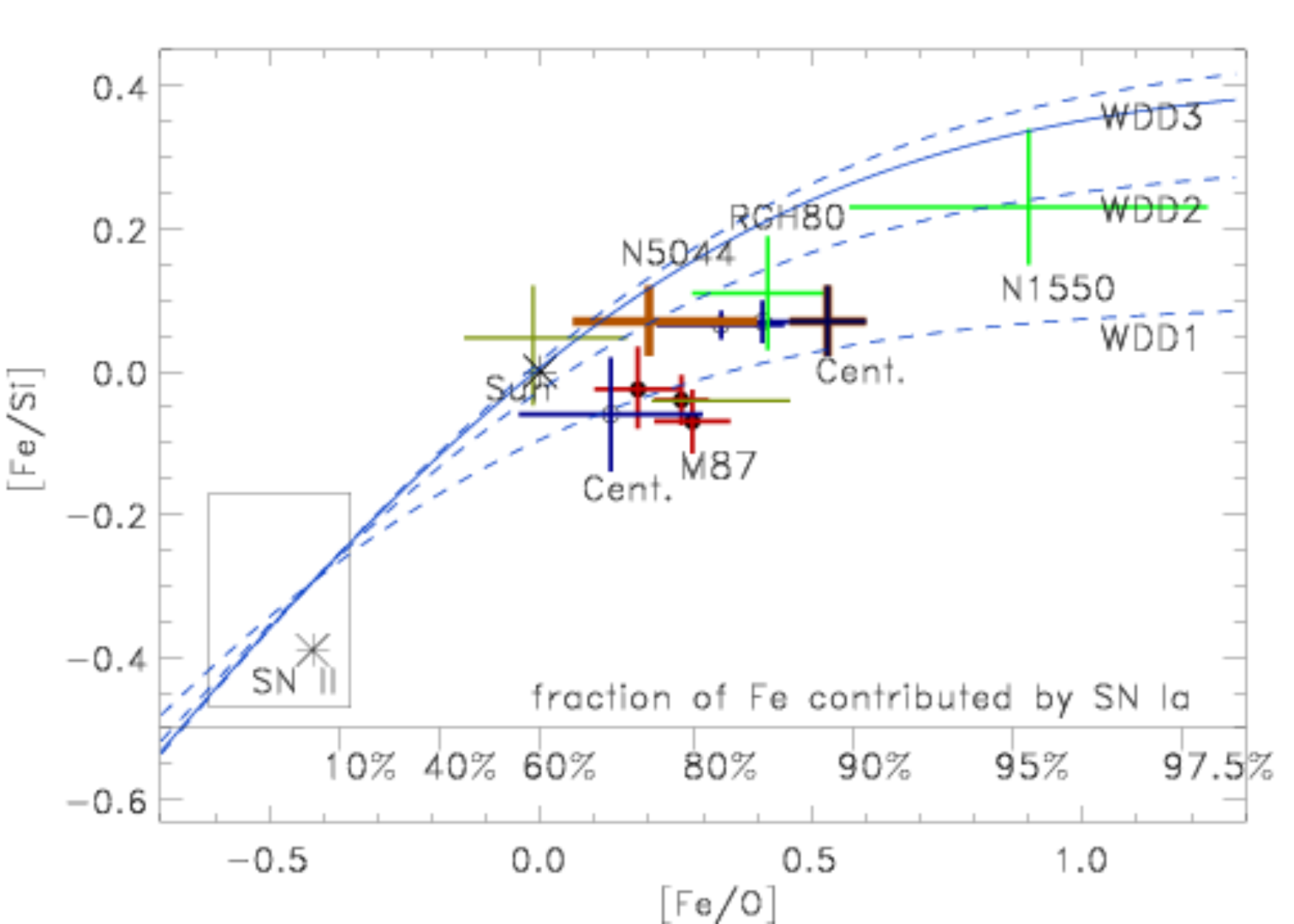}}
\caption{
\footnotesize
Comparison of the Fe/O and Fe/Si abundance ratios for different 
radial regions in different galaxy clusters with data taken from 
(\citep{mat03} M87; red), (\citep{mat04} Centaurus, blue), 
(\citep{sun03} NGC1550, green),
(\citep{xue04}  RGH80, green), (\citep{buo03} NGC5044, orange thick), 
(\citep{tam04} several cool clusters observed with XMM, light green 
thin), and (\citep{dup00} A496
observed with ASCA, blue hidden behind N5044 data points). 
The solid line gives the W7 model \citep{nom84}  which 
accounts for the abundance of our sun. The three dashed lines give the 
deflagration-detonation models WDD1, WDD2, and WDD3 \citep{iwa99}.
The lower asterisk indicates the SN II yields, the upper asterisk the composition
of the Sun. The box indicates the composition of low metallicity halo stars in
our galaxy \citep{cle99}.
The range of observational values is well bracketed by the 
different theoretical SN Ia yields.
}
\label{fig9}
\end{figure}

One can further study the possibility of two types of SN Ia contributing 
to the ICM enrichment with the diagram shown in Fig.\ref{fig9} \citep{boe05}.
Since O is hardly produced by SN Ia, one can start with the Fe/O ratio
yields of SN cc and trace the additional contribution of SN Ia to
the enrichment by an increasing Fe/O ratio. This parameter is plotted at 
the X-axis of the Figure. We can do the same for the Fe/Si ratio
and plot it on the Y-axis. A certain SN Ia model will then define a curve
in the plane in the figure. We have plotted several curves
for the classical W7 model \citep{nom84} and for different deflagration-detonation
models tracing different trajectories in the plot. We also plotted
a series of observations for clusters and groups and different radii in the
clusters. The central regions and in particular the central regions of groups 
which are very much dominated by the central elliptical galaxy appear on
the right of the plot with the largest relative contribution by
SN Ia. The fraction of the Fe mass contributed by SN Ia is labled at
the bottom of the plot.

We also note in the figure, that the observed abundance ratios fall in between 
the classical model W7 and the various DD models. It is still
too early to draw a firm conclusion. But there are signs that the old stellar
population contributing to the enrichment of the central ICM in cool core
clusters has a lower Fe/Si ratio than our Galaxy. This may imply that the SN Ia 
yields in old stellar populations have a higher Si/Fe ratio than that
of younger populations and that there are different types of supernovae of
type Ia.

Studies on the change of the ICM metallicity as a function of redshift
are still in their infancy and different publications find somewhat
different trends. Whereas Balestra et al. (2007) and Maughan et al. (2008)
find a clearer trend of a decreasing metallicity with redshift up
to $z = 1$ and beyond, a recent study by Baldi et al. (2012) which 
separates the central and the outer regions of the ICM cannot establish
a significant abundance evolution. But a decreasing metallicity trend is
also seen in their data.

\section{Conclusions}

The heavy element abundances in the ICM of clusters of galaxies provide
a lot of interesting and important information on the history of the
nucleosysthesis of these metals, on their sources in the clusters and 
on the possible transport processes. They also allow us to test
the supernova models through their predicted abundance yields.
But better observational data are needed.
Significant progress can be expected, when the new X-ray 
calorimeter detectors become available with much better spectral resolution
and well resolved lines. The first space mission
that can provide this capability is ASTRO-H which will hopefully
be launch successfully soon.
 
\begin{acknowledgements}
I like to thank the organizers of the meeting for an
interesting and stimulating conference. 
\end{acknowledgements}

\bibliographystyle{aa}

\begin{thebibliography}{}

\bibitem[{Allen et al. (2011)}]{all12}
Allen, S.W., Evrard, A.E., Mantz, A.B., 2011,
ARA\&A, 49, 409

\bibitem[{Anders \& Grevesse (1989)}]{and89}
Anders, E. \& Grevesse, N., 1989, GeCoA, 53, 197

\bibitem[{Asplund et al. (2009)}]{asp09}
Asplund, M., Grevesse, N., Sauval, A. J., et al. 2009, ARA\&A, 47, 481

\bibitem[{Baldi et al. (2012)}]{bal12}
Baldi, A., Ettori, S., Molendi, S., et al.  2012, A\&A, 537, 142 

\bibitem[{Balestra  et al. (2007)}]{bal07}
Balestra, I., Tozzi, P., Ettori, S., et al., 2007, A\&A, 462, 429

\bibitem[{B\"ohringer \& Hensler (1989)}]{boe89}
B\"ohringer, H. \& Hensler, G., 1989,
A\&A, 215, 147 

\bibitem [{B\"ohringer et al. (2001)}]{boe01}
B\"ohringer, H., Belsole, E., Kennea, J., et al., 2001
A\&A, 365, L181 

\bibitem[{B\"ohringer et al. (2004)}]{boe04}
B\"ohringer, H., Matsushita, K., Churazov, E., et al., 2004,
A\&A, 416, L21

\bibitem[{B\"ohringer et al. (2005)}]{boe05}
B\"ohringer H., Matsushita, K., Finoguenov, A., et al., 2005.
AdSpR, 36, 677

\bibitem[{B\"ohringer  (2008)}]{boe10}
B\"ohringer, H., in {\it The Universe in X-rays}, Tr\"umper, J.E.,
Hasinger, G., eds., Springer 2008

\bibitem[{B\"ohringer \& Werner  (2010)}]{bow10}
B\"ohringer, H. \& Werner, N., A\&AR, 18 127 


\bibitem[{Buote et al.  (2003)}]{buo03}
Buote, D.A., Lewis, A.D., Brighenti, F., et al.,
2003, ApJ, 595, 151


\bibitem[{Clementini et al. (1999)}]{cle99}
Clementini, G., Gratton, R.G., Carretta, E., et al., 1999, 
MNRAS, 302, 22 


\bibitem[{De Grandi \& Molendi (2001)}]{dgr01}
De Grandi, S. \& Molendi, S., 2001, 
ApJ, 551, 153

\bibitem[{De Grandi et al. (2004)}]{dgr04}
De Grandi, S., Ettori, S., Longhetti, M., et al., 2004, 
A\&A, 419, 7 

\bibitem[{de Plaa et al. (2007)}]{dep07}
de Plaa, J., Werner, N., Bleeker, J.A.M., et al., 
2007, A\&A, 465, 345

\bibitem[{de Plaa (2013)}]{dep12}
de Plaa, J., 2013 AN, 334, 416	
	
\bibitem[{Dupke \& White (2000)}]{dup00}
Dupke, R.A.  \& White, R.E.,
2000, ApJ, 528, 139

\bibitem[{Finoguenov et al. (2002)}]{fin02}
Finoguenov, A., Matsushita, K., B\"ohringer, H., et al., 2002, 
A\&A, 381, 21

\bibitem[{Iwamoto et al. (1999)}]{iwa99}
Iwamoto, K., Brachwitz, F., Nomoto, K., et al., 1999, 
ApJ SS, 125, 439 

\bibitem[{Lodders (2003)}]{lod03}
Lodders, K., 2003 ApJ, 591, 1220


\bibitem[{Matsushita et al. (2003)}]{mat03}
Matsushita, K., Finoguenov, A., B\"ohringer, H., 2003, 
A\&A, 401, 443 

\bibitem[{Matsushita et al. (2007)}]{mat04}
Matsushita, K., B\"ohringer, H., Takahashi, I., et al., 2007,
A\&A, 462, 953

\bibitem[{Maughan et al. (2008)}]{mau08}
Maughan, B.J., Jones, C., Forman, W., et al., 2008,
ApJS, 174, 117

\bibitem[{Mewe et al. (1995)}]{mew95}
Mewe, R., Kaastra, J.S., Liedahl, D.A., Legacy, 6, 16

\bibitem[{Navarro et al. (1995)}]{nfw95}
Navarro, J.F., Frenk, C.S., White, S.D.M., 1995, MNRAS, 275, 720

\bibitem[{Nomoto et al. (1984)}]{nom84}
Nomoto, K., Thielemann, F.-K., Yokoi, K., 1984, 
ApJ, 286, 644


\bibitem[{Rebusco et al. (2006)}]{reb06}
Rebusco, P., Churazov, E., B\"ohringer, H., et al., 2006,
MNRAS, 372, 1840

\bibitem[{Sanders \& Fabian (2006)}]{san06}
Sanders, J.S. \& Fabian, A.C., 2006, MNRAS, 371, 1483

\bibitem[{Sarazin (1986)}]{sar86}
Sarazin, C.L., 1986, RvMP, 58, 1

\bibitem[{Smith et al. (2001)}]{smi01}
Smith, R.K., Brickhouse, N.S., Liedahl, D.A. et al., 2001, ApJ, 556, L91


\bibitem[{Sun et al. (2003)}]{sun03}
Sun, M., Forman, W., Vikhlinin, A., et al., 2003, 
ApJ, 598, 250



\bibitem [{Tamura et al. (2004)}]{tam04}
Tamura, T., Kaastra, J.S., den Herder, J.W.A., et al., 2004, 
A\&A, 420, 135



\bibitem[{Tamura et al. (2009)}]{tam09}
Tamura, T., Maeda, Y., Mitsuda, K., et al., 2009, ApJ, 705, L62
 
\bibitem[{Werner et al. (2006)}]{wer06}
Werner, N., B\"ohringer, H., Kaastra, J.S., et al., 2006, A\&A, 459, 353 

\bibitem[{Xue et al. (2004)}]{xue04}
Xue, Y.-J., B\"ohringer, H., Matsushita, K, 2004, 
A\&A, 420, 833








\end{thebibliography}

\end{document}